\documentclass[12pt,twoside]{article}
\usepackage[mathscr]{eucal}
\usepackage{amsmath,amsfonts,amssymb,amsthm}
\usepackage{hyperref}
\usepackage[matrix,arrow]{xy}
\usepackage{times}
\usepackage{graphicx}
\usepackage{epstopdf}
\usepackage{multibox}
\usepackage{dcolumn}

\def\d_Vphi{\text{d}_V\hspace{-0.06em}\phi}
\def\d_Vphibar{\text{d}_V\hspace{-0.06em}\bar\phi}
\def\d_Vxi{\text{d}_V\hspace{-0.06em}\xi}

\def\be{\begin{eqnarray}}
\def\ee{\end{eqnarray}}
\def\beann{\begin{eqnarray*}}
\def\eeann{\end{eqnarray*}}
\def\beq{\begin{equation}}
\def\eeq{\end{equation}}
\def\ba{\begin{array}}
\def\ea{\end{array}}
\def\ben{\begin{enumerate}}
\def\een{\end{enumerate}}
\def\bea{\begin{eqnarray}}
\def\eea{\end{eqnarray}}

\def\5{\bar }
\def\6{\partial }
\def\7{\hat }
\def\4{\tilde }
\advance\textheight\topskip
\renewcommand{\tilde}{\widetilde}
\renewcommand{\hat}{\widehat}

\newcommand{\bref}[1]{\textbf{\ref{#1}}}

\renewcommand{\d}{\partial}

\renewcommand{\geq}{\,{\geqslant}\,}

\newcommand{\binner}[2]{%
  {\langle}\kern-4.15pt{\langle}#1{,}\,#2{\rangle}\kern-4.15pt{\rangle}}

\newcommand{\half}{\frac{1}{2}}

\newcommand{\ffrac}[2]{\raisebox{.5pt}%
  {\footnotesize$\displaystyle\frac{#1}{#2}$}\kern1pt}

\def\cH{\mathcal{H}}

\def\cO{\mathcal{O}}

\def\cR{\mathcal{R}}

\numberwithin{equation}{section} \makeatletter
\@addtoreset{equation}{section}

\begin{document}

\pagestyle{myheadings} \markboth{\textsc{\small Barnich, Troessaert
}}{\textsc{\small Duality and integrability}} \addtolength{\headsep}{4pt}

\begin{flushright}\small
ULB-TH/08-42
\end{flushright}

\begin{centering}

  \vspace{1cm}

  \textbf{\large{Duality and integrability: Electromagnetism,
      linearized gravity and massless higher spin gauge fields as
      bi-Hamiltonian systems}}

 \vspace{1.5cm}

 {\large Glenn Barnich$^{a}$ and C\'edric Troessaert$^{b}$}


 \vspace{.5cm}

 \begin{minipage}{.9\textwidth}\small \it \begin{center}
     Physique Th\'eorique et Math\'ematique, Universit\'e Libre de
     Bruxelles\\ and \\ International Solvay Institutes, \\ Campus
     Plaine C.P. 231, B-1050 Bruxelles, Belgium \end{center}
 \end{minipage}

 \vspace{1cm}

 \end{centering}

 \vspace{1cm}

 \begin{center}
   \begin{minipage}{.9\textwidth}
     \textsc{Abstract}.  In the reduced phase space of
     electromagnetism, the generator of duality rotations in the usual
     Poisson bracket is shown to generate Maxwell's equations in a
     second, much simpler Poisson bracket. This gives rise to a
     hierarchy of bi-Hamiltonian evolution equations in the standard
     way. The result can be extended to linearized Yang-Mills theory,
     linearized gravity and massless higher spin gauge fields.
   \end{minipage}
 \end{center}

 \vfill

 \noindent
 \mbox{}
 \raisebox{-3\baselineskip}{%
   \parbox{\textwidth}{\mbox{}\hrulefill\\[-4pt]}}
 {\scriptsize$^a$Research Director of the Fund for Scientific
   Research-FNRS (Belgium).\\ $^{b}$Research Fellow of the Fund for
   Scientific Research-FNRS (Belgium).}

\thispagestyle{empty}
\newpage



\section{Introduction}
\label{sec:introduction}

A cornerstone of soliton theory is the discovery that the evolution
equations are Hamiltonian systems
\cite{gardner:1548,Zakharov:1971qm}. In this context, the occurrence of
hierarchies of evolution equations sharing the same infinite set of
conservation laws can be understood as a consequence of the existence
of a second, compatible Hamiltonian structure giving rise to the same
evolution equations \cite{magri:1156,Gelfand:1979zh}.

The equations of motion associated to the theories for the known fundamental
forces of nature, electromagnetism, Yang-Mills theories and
gravitation, are variational and thus Hamiltonian. This is no
coincidence, since these theories are fundamentally quantum, at least
the first three of them, and only for variational theories
quantization is sufficiently well understood. 

In order to have Poincar\'e invariance, respectively diffeomorphism
invariance, manifestly realized, most modern investigations of these
equations are carried out in the Lagrangian framework. This could be
the reason why the bi-Hamiltonian structure underlying these equations
and discussed below has hitherto remained unnoticed.

At the heart of our analysis is an important exception to this
paradigm, namely the question whether the duality invariance of the
four dimensional Maxwell or linearized gravity equations
admits a canonical generator. This question has been answered to the
affirmative in the reduced phase space of these theories and
generalized to massless higher spin gauge fields
\cite{Deser:1976iy,Henneaux:2004jw,Deser:2004xt}.

We will show in this letter that the reduced phase space formulation
of massless higher spin gauge fields is bi-Hamiltonian. The second
Poisson bracket on reduced phase space turns out to be more natural
than the one induced from the covariant variational principle, while
the generator for duality rotations plays the role of the second
Hamiltonian.

We will start with the most familiar case of Maxwell's equations, for
which all details will be given. This result trivially generalizes to
Yang-Mills theory with an invariant, non degenerate metric, linearized
around a zero potential by decorating the expressions obtained in the
electromagnetic case with an additional Lie algebra index.

The reduced phase space of linearized gravity around flat spacetime
consists of two symmetric transverse and traceless potentials
\cite{Arnowitt:1962aa}. The entire analysis of the electromagnetic
case then carries over in a straightforward way by taking care of the
additional spatial index. The same goes for massless gauge fields of
spin higher than $2$.

Several generalizations and extensions are suggested by the result
reported here. A first exercise consists in studying the consequences
for symmetries and conservation laws of both the Maxwell and the
higher spin equations and compare them to known results (see
e.g.~\cite{0305-4470-25-5-004,2003,pohjanpelto:2007} and references
therein). Another obvious question is to investigate more general
backgrounds. For instance, the generalization to massless spins
propagating on (anti-) de Sitter spaces instead of Minkowski spacetime
and the inclusion of fermionic gauge fields should be
straightforward. In Yang-Mills theories (anti) self-dual backgrounds could
be promising in view of their close connection to integrable systems.

The most important problem is however the inclusion of
interactions. When comparing to the Korteweg-de Vries equation for
instance, the present work corresponds to the bi-Hamiltonian structure
for the linearized equation. The question is then to find interactions
that preserve this structure. This will most probably not work for the
standard interactions, as it is known that duality invariance does not
survive the standard Yang-Mills type coupling, nor the extension from
linearized to full gravity \cite{Deser:1976iy,Deser:2005sz}. A
promising candidate for an interacting model in which duality
invariance does survive is Born-Infeld theory\footnote{The authors
  thank X. Bekaert for pointing this out.}. We plan to address at
least some of these questions elsewhere.

\section{Electromagnetism}
\label{sec:electromagnetism}

The first order Hamiltonian action for free electromagnetism is
\begin{gather}
  \label{eq:1}
  S[A_i,E^i,A_0]=\int dt \Big[\int d^3x\ (-E^i \d_0  A_i -A_0 \d_i E^i)-
  H\Big],\\  H=\half \int d^3x\ [E^i E_i + (\cO A)^i(\cO A)_i].
\end{gather}
Here, $A_i$ is the vector potential, $E^i$ the electric field and $A_0$
the Lagrange multiplier for the Gauss constraint. We use $x^0=t$ and
$\cO$ is the curl, $(\cO A)^i=\epsilon^{ijk}\d_j A_k$. Unless
otherwise specified, the summation convention is used, Roman indices
$i,j,k\dots$ take values from $1$ to $3$, are raised and lowered with
the Kronecker delta, while $\epsilon_{ijk}$ is completely
skew-symmetric with $\epsilon_{123}=1$.

The decomposition of $A_i,E^i$ into transverse and longitudinal
vectors $A_i=A_i^T+A_i^L,E^i=E^{Ti}+E^{Li}$ allows one to eliminate the
pure gauge degrees of freedom $A_i^L,E^{iL}$ associated with the Gauss
constraint $\d_i E^i=0$. By introducing a second vector potential $Z_i$
such that $E^{iT}=(\cO Z)^i$ the reduced phase space action has been
shown to be invariant under duality rotations
\cite{Deser:1976iy}. This is most transparent in the double potential
notation \cite{Schwarz:1993vs} where $A_i^a\equiv (A_i,Z_i)$. Roman
indices $a,b,c\dots$ take values from $1$ to $2$, are raised and
lowered with the Kronecker delta, while $\epsilon_{ab}$ is completely
skew-symmetric with $\epsilon_{12}=1$. In this notation, the
reduced phase space action can be written as 
\begin{gather}
  \label{eq:2}
  S^R[A^{Ta}_i]= \int dt \Big[\int d^3x\ \half \epsilon_{ab} (\cO A^{Ta})^i
 \d_0  A^{Tb}_i-  H_1\Big],\\
H_1=\half \int d^3x\ (\cO A^{Ta})_i(\cO A^T_{a})^i=-\half \int d^3x\
A^{Tai}\Delta A^T_{ai},
\end{gather}
where in the second expression for the Hamiltonian, we have used that
$\cO$ is ``self-adjoint'', 
\begin{equation}
\int d^3x\ g_i (\cO f)^i=\int d^3 x\ (\cO
g)_i f^i,\label{eq:21}
\end{equation}
and 
\begin{equation}
(\cO^2 f^T)^i=-\Delta f^{Ti}\label{eq:22},
\end{equation}
with $\Delta=\d_i \d^i$ the Laplacian in flat space. The standard
Poisson bracket determined by the kinetic term is
\begin{equation}
  \label{eq:3}
  \{A^{Ta}_i(x),A^{Tbj}(y)\}_1=\epsilon^{ab}\Delta^{-1}
\epsilon^{jkl}\d^y_k\delta^{T(3)}_{il}(x-y)=\epsilon^{ab}\Delta^{-1}\big(
  \cO^y \delta^{T(3)}_{i}(x-y)\big)^j,
\end{equation}
where $\delta^{T(3)}_{ij}(x-y)$ is the transverse delta function, see
e.g.~\cite{cohentannoudji:1989} section $A_1.2$. In vacuum, Maxwell's
equations for the physical degrees of freedom read
\begin{equation}
  \label{eq:4}
  \d_0 A^{Tai}(x)=\{A^{Tai}(x),H_1\}_1=-\epsilon^{ab}(\cO A^T_{b})^i(x),
\end{equation}
while the generator for duality rotations is 
\begin{equation}
  \label{eq:5}
  H_0=-\half \int d^3x\  A^{ai}_T(\cO A_{a}^T)_i, \qquad
  \{H_0,H_1\}_1=0.  
\end{equation}
Remarkably, the duality generator is simply an $SO(2)$ Chern-Simons
action \cite{Deser:1997xu}.

When presented in this way, the second Hamiltonian structure is
obvious and a lot simpler than the one induced from the covariant
action principle. Indeed, a natural Poisson bracket on reduced
phase space is simply
\begin{equation}
  \label{eq:6}
  \{A^{Ta}_i(x),A^{Tb}_j(y)\}_0=\epsilon^{ab}\delta^{T(3)}_{ij}(x-y),
\end{equation}
in terms of which the duality generator is the new Hamiltonian for
Maxwell's equations,
\begin{equation}
  \label{eq:7}
  \{A^{Tai}(x),H_1\}_1=\{A^{Tai}(x),H_0\}_0.
\end{equation}
This is our main result. 

At this stage, one can pause and ask whether electromagnetism and its
quantization should not be based on this new Hamiltonian structure. A good
reason to favour the old, more complicated structure is that, by
construction, the Poincar\'e and conformal symmetries admit canonical
generators for the old structure, while not all of them do for the new
one. We plan to return to this question in detail elsewhere. 

The reason why we are interested in the new structure is not so much
in order to use it as an alternative for quantization, but because of
what it tells us about Maxwell's equations themselves. For instance,
using the bi-Hamiltonian structure, there is a standard way to
generate an infinite set of generators that commute with the
Hamiltonian and are in involution, which we now briefly describe (see
e.g.~\cite{Olver:1993}, chapter 7.3 for an elementary introduction).

The recursion operator is obtained by contracting the new Poisson
structure with the inverse of the original one,
\begin{equation}
  \label{eq:8}
  {\cR^{ai}}_{bj}=-\delta^a_b(\cO)^i_j.
\end{equation}
Consider, for $p\geq 1$, $K^{Tai}_{p}=(-)^p\epsilon^{ab}(\cO^{p}
A_b^T)^i$, or equivalently,
\begin{equation}
  \label{eq:10}
  K^{Tai}_{2n+1}(x)=(-)^{n+1}\epsilon^{ab}\Delta^n(\cO A_b^T)^i(x),\quad
  K^{Tai}_{2n+2}(x)=(-)^{n+1}\epsilon^{ab}\Delta^{n+1}A_b^{Ti}(x),
\end{equation}
for $n\geq 0$. 
The evolution equations of the hierarchy 
\begin{equation}
  \label{eq:9}
  \d_0 A^{Tai}(x)=K^{Tai}_p(x),\quad \forall p\geq 1,
\end{equation}
are also bi-Hamiltonian, 
\begin{equation}
  \label{eq:11}
  K^{Tai}_p(x)=\{A^{Tai}(x),H_p\}_1=\{A^{Tai}(x),H_{p-1}\}_0,
\end{equation}
where $H_{p-1}=\frac{(-)^p}{2}\int d^3x\ A^{Ta}_i(\cO^p A^T)_i$, 
\begin{equation}
  \label{eq:13}
  H_{2n}=\frac{(-)^{n+1}}{2} \int d^3x\  A^{Tai}\Delta^n (\cO
  A_{a}^T)_i,\
  H_{2n+1}=\frac{(-)^{n+1}}{2} \int d^3x\ A^{Ta}_i\Delta^{n+1} A^{T}_{ai}.
\end{equation}
with Hamiltonians that are in involution, 
\begin{equation}
  \label{eq:12}
  \{H_n,H_m\}_1=0=\{H_n,H_m\}_0,\quad \forall n,m\geq 0. 
\end{equation}

\section{Linearized gravity}
\label{sec:linearized-gravity}

The Hamiltonian formulation of general relativity linearized around
flat spacetime is based on the first order
action principle \cite{Arnowitt:1962aa}
\begin{gather}
  \label{eq:2f}
  S_{PF}[h_{mn},\pi^{mn},n_m,n]=\int dt\Big[\int
  d^3x\, \big(\pi^{mn}\dot h_{mn}-n^m\cH_m-n\cH \big)-H_{PF}\Big], 
\\
H_{PF}[h_{mn},\pi^{mn}]=\int
d^3x\big(\pi^{mn}\pi_{mn}-\half\pi^2+\frac{1}{4}\d^rh^{mn}\d_rh_{mn}-\nonumber\\
-\half\d_m h^{mn}\d^r h_{rn}+\half\d^m h\d^n h_{mn}-\frac{1}{4}\d^m
h\d_m h\big),\label{H}\\
\cH_m=-2\d^n\pi_{mn},\quad
\cH_\perp=\Delta h-\d^m\d^n h_{mn}.\label{eq:2a}
\end{gather}
Here, $h={h^m}_m$, $\pi={\pi^m}_m$ and the linearized $4$ metric is
reconstructed using $h_{00}=-2n$ and $h_{0i}= n_i$.

Duality invariance of massless spin $2$ fields has been uncovered by
introducing suitable potentials for the linearized three metric and
its momentum \cite{Henneaux:2004jw}. We will use below properties of
these potentials discussed in more details
in~\cite{deser:1967aa,Deser:2004xt,Barnich:2008ts}.

Symmetric rank two tensors $\phi_{mn}$ decompose as
\begin{eqnarray}
\phi_{mn} & = & \phi^{TT}_{mn} + \phi^T_{mn} + \phi^L_{mn},\\
\phi^L_{mn} & = & \partial_m \psi_n + \partial_n \psi_m,\\
\phi^T_{mn} & = & \half \left( \delta_{mn} \Delta-\partial_m \partial_n
\right) 
\psi^T.\label{eq:dec}
\end{eqnarray}
The tensor $\phi^T_{mn}$ contains the trace of the transverse part of
$\phi_{mn}$ and only one independent component. Three independent
components are encoded in the longitudinal part $\phi^L_{mn}$, 
while the transverse-traceless part $\phi^{TT}_{mn}$ containing two
independent components is defined as the remainder, 
\begin{eqnarray}
\phi^{TT}_{mn} & = & \phi_{mn} - \phi^T_{mn} - \phi^L_{mn}.
\end{eqnarray}
The elements of the decomposition are orthogonal under 
integration if boundary terms can be neglected,
\begin{equation}
  \int d^3 x\, \phi^{mn} \varphi_{mn} = \int d^3 x 
  \left( \phi^{TTmn} \varphi^{TT}_{mn} +\phi^{Lmn} \varphi^L_{mn} + 
\phi^{Tmn} \varphi^T_{mn} \right).
\end{equation} 

The generalized curl
\cite{Deser:2004xt,Deser:2005sz} is defined through,
\begin{eqnarray}
  \label{eq:12a}
  \left(\cO\phi\right)_{mn}=\half(\epsilon_{mpq} \partial^p {\phi^{q}}_n + 
\epsilon_{npq} \partial^p  {\phi^{q}}_m).
\end{eqnarray}
It satisfies 
\begin{equation}
  \label{eq:17}
  \left(\cO(\cO\phi^{TT})\right)_{mn}=-\Delta \phi^{TT}_{mn}. 
\end{equation}
and is self-adjoint,
\begin{eqnarray}
  \label{eq:7a}
  \int d^3 x\, \left( \cO \phi^{TT}\right)^{mn}\varphi^{TT}_{mn}=
\int d^3 x\, \phi^{TT mn}\left( \cO \varphi^{TT}\right)_{mn}.
\end{eqnarray}
The reduced phase space action for linearized gravity is 
\begin{gather}
  \label{eq:18}
  S^R=\int dt\Big[ \int d^3x\ \pi^{TTmn}\d_0 h^{TT}_{mn}-H_1\Big],\\
H_1=\int d^3x\ \Big(
  \pi^{mn}_{TT}\pi_{mn}^{TT}+\frac{1}{4}\d_rh^{TT}_{mn}\d^rh_{TT}^{mn}\Big). 
\end{gather}

In the reduced phase space, one can make the
change of variables, 
\begin{equation}
  \label{eq:16}
  h^{TT}_{mn} =  2\left(\cO H^{1TT}\right)_{mn}, \qquad
\pi^{TT}_{mn} = -\Delta H^{2TT}_{mn}.
\end{equation}
so that the reduced phase space action becomes 
\begin{gather}
  \label{eq:19}
S^R[H^{TT a}_{mn}]=  \int dt\Big[ \int d^3x\  \epsilon_{ab} \Delta\left( \cO
  H^{TTa}\right)^{mn} 
\d_0 H^{TTb}_{mn}-H_1\Big],\\
H_1= \int d^3x\ \Big( H^{TTamn} \Delta^2  H^{TT}_{amn}).
\end{gather}
The standard Poisson bracket determined by the kinetic term is 
\begin{equation}
  \label{eq:23}
  \{H^{TT a}_{mn}(x),H^{TT b
    kl}(y)\}_1=\half\epsilon^{ab}\Delta^{-2}\left(\cO^y \delta^{(3)
      TT}_{mn}(x-y)\right)^{kl},
\end{equation}
where $\delta^{(3) TTkl}_{mn}(x-y)$ denotes the projector on the
transverse-traceless part of a symmetric rank two tensor. The duality
generator is 
\begin{equation}
  \label{eq:20}
  D=- \int d^3x\ H^{TT amn}\Delta\left(\cO H^{TT}_a\right)_{mn},\quad
  \{D,H_1\}_1=0. 
\end{equation}

The analogy with the spin $1$ case can be made perfect by the change
of variables,
\begin{equation}
  \label{eq:24}
  H^{1TT}_{mn}=\frac{1}{\sqrt{2}} \Delta^{-1}\left(\cO
    A^{2TT}\right)_{mn},\qquad H^{2TT}_{mn}=\frac{1}{\sqrt{2}} 
\Delta^{-1}\left(\cO  A^{1TT}\right)_{mn}. 
\end{equation}
in terms of which 
\begin{gather}
  \label{eq:25}
  H_1= \half \int d^3x\ \Big( \left(\cO A^{TTa}\right)^{mn}  \left(
   \cO A^{TT}_a\right)_{mn}\Big)=-\half \int d^3x\ \Big(
 A^{TTamn}\Delta  A^{TT}_{amn}\Big),\\
\{A^{TT a}_{mn}(x),A^{TT b
    kl}(y)\}_1=\epsilon^{ab}\Delta^{-1}\left(\cO^y \delta^{(3)
      TT}_{mn}(x-y)\right)^{kl},\\
H_0=-D= -\half \int d^3x\ A^{TT amn}\left(\cO A^{TT}_a\right)_{mn}. 
\end{gather}

All formulae of section~\bref{sec:electromagnetism} below
equation~\eqref{eq:5} now generalize in a straightforward way to
massless spin $2$ fields by replacing $T$ (transverse) by $TT$
(transverse-traceless) and contracting over the additional spatial
index.

\section{Massless higher spin gauge fields}
\label{sec:higher-spin-gauge}

The extension of these results to massless higher spin gauge fields
\cite{Fronsdal:1978rb} (see also \cite{deWit:1979pe}) follows directly
from the observation that the Hamiltonian reduced phase space
formulation of these theories merely involves additional spatial
indices \cite{Deser:2004xt}, so that all above results generalize in a
straightforward way.

This can be seen for instance by starting from the approach inspired
from string field theory, where the Lagrangian action for massless
higher spin gauge fields is written as the mean value of the BRST
charge for a suitable first quantized particle model
\cite{Ouvry:1986dv,Bengtsson:1986ys,Henneaux:1987cp} (see also
\cite{Sagnotti:2003qa,Barnich:2004cr} for further developments). In
this framework the reduction of the action to the light-cone gauge
corresponds to the elimination of BRST quartets composed of ghost and
light-cone oscillators
(see~e.g.\cite{Siegel:1999ew,Barnich:2005ga}). In exactly the same
way, the ghost, temporal and longitudinal oscillators form quartets
that can be eliminated to yield the Lagrangian gauge fixed action for
a massless field of spin $s\geq 1$,
\begin{equation}
  \label{eq:14}
  S_L[\phi^{TT}_{i_1\dots i_s}]=-\half \int d^4x\ \d_\mu
\phi^{TT}_{i_1\dots i_s}\d^\mu \phi^{TTi_1\dots i_s}
\end{equation}
where the field $\phi^{TT}_{i_1\dots i_s}$ is real, completely symmetric,
traceless and transverse,
\begin{equation}
  \label{eq:15}
  \phi^{TT}_{i_1\dots i_s}=\phi^{TT}_{(i_1\dots i_s)},\  \phi^{TT i}_{i
    i_3\dots i_s}=0,\ \d^i  \phi^{TT}_{i i_2\dots i_s}=0. 
\end{equation}
The Hamiltonian formulation is direct, the momenta being
$\pi^{TT}_{i_1\dots i_s}=\d_0 \phi^{TT}_{i_1\dots i_s}$. 

Consider then the Fock space defined by $[a^i,
a^{\dagger}_j]=\delta^i_j$, $a_i|0\rangle=0$, the number operator
$N=a^\dagger_i a^i$, the ``string field''
$\phi^{TT}_s(x)=\frac{1}{\sqrt{s!}}a^\dagger_{i_1}\dots
  a^\dagger_{i_s}|0\rangle \phi^{TT}_{i_1\dots i_s}(x)$ and the inner product
\begin{equation}
    \label{eq:29}
    \langle \phi^{TT}_s,\psi^{TT}_s\rangle=\int d^3x\ \langle
    \phi^{TT}_s,\psi^{TT}_s\rangle_F=\int d^3x\,   \phi^{TT}_{i_1\dots
      i_s} \psi^{TT i_1\dots i_s}. 
\end{equation}
With this inner product, the generalized curl \cite{Deser:2004xt} 
\begin{equation}
\cO =\frac{1}{N}\epsilon^{ijk}
a^\dagger_i\d_j a_k\label{eq:30}
\end{equation}
is again self-adjoint. Furthermore, it squares to $-\Delta$
inside the inner product involving transverse-traceless fields, 
\begin{multline}
  \label{eq:31}
  \cO^2=\frac{1}{N^2}\big[-\Delta N^2+(\partial\cdot
  a^\dagger)(\partial\cdot a) +(a^\dagger\cdot a^\dagger)\Delta(a\cdot
  a)+ 2(\partial\cdot a^\dagger) N (\partial\cdot a)\\-(\partial \cdot
  a^\dagger)^2 (a\cdot a) -(a^\dagger\cdot a^\dagger)(\partial\cdot
  a)^2\big]
\Longrightarrow  \langle \phi^{TT}_s,\cO^2\psi^{TT}_s\rangle=-
 \langle \phi^{TT}_s,\Delta\psi^{TT}_s\rangle.
\end{multline}

The change of variables making duality invariance transparent is
\begin{equation}
\phi^{TT}_{i_1\dots i_s}=A^{TT 1}_{i_1\dots i_s},\ \pi^{TT}_{i_1\dots
  i_s}=-\left(\cO A^{TT2}\right)_{i_1\dots i_s}\label{eq:26}. 
\end{equation}
The first order reduced phase space action becomes
\begin{gather}
  \label{eq:28}
  S^R[A^{TT a}]=\int dt\Big[\int d^3x\ \half \epsilon_{ab} \left(\cO
    A^{TTa}\right)^{i_1\dots i_s} \d_0A^{TT b}_{i_1\dots
    i_s}-H_1\Big],\\
H_1=-\half \int d^3x\ A^{TT a}_{i_1\dots i_s}\Delta A^{TTi_1\dots i_s}_a. 
\end{gather}
Again, all formulae of section~\bref{sec:electromagnetism} below
equation~\eqref{eq:4}, including the one for the duality generator,
suitably generalize by contracting over the additional spatial
indices.

\section*{Acknowledgements}
\label{sec:acknowledgements}

\addcontentsline{toc}{section}{Acknowledgments}

The authors thank M.~Henneaux for a most useful discussion. This work
is supported in part by a ``P{\^o}le d'Attraction Interuniversitaire''
(Belgium), by IISN-Belgium, convention 4.4505.86, by the Fund for
Scientific Research-FNRS (Belgium), and by the European Commission
programme MRTN-CT-2004-005104, in which the authors are associated to
V.U.~Brussel.



\def\cprime{$'$}
\providecommand{\href}[2]{#2}\begingroup\raggedright\endgroup

\end{document}